\documentclass[twocolumn,aps,showpacs,showkeys]{revtex4}
\usepackage{graphicx}
\usepackage{epsfig}
\usepackage{amsmath}

\renewcommand{\Re}{\mathrm{Re}}

\begin{document}

\title{Control of unstable steady states by extended time-delayed feedback}
\author{Thomas Dahms}
\author{Philipp H{\"o}vel}
\author{Eckehard Sch{\"o}ll}\email{schoell@physik.tu-berlin.de}

\affiliation{Institut f{\"u}r Theoretische Physik, Technische
Universit{\"a}t Berlin, 10623 Berlin, Germany}
 
\date{\today}

\begin{abstract}
Time-delayed feedback methods can be used to control unstable periodic orbits as well as unstable steady states. We
present an application of extended time delay autosynchronization introduced by Socolar et al. to an unstable focus.
This system represents a generic model of an unstable steady state which can be found for instance in a Hopf
bifurcation. In addition to the original controller design, we investigate effects of control loop latency and a bandpass filter on the domain of control. Furthermore, we consider coupling of the control force to the system via a
rotational coupling matrix parametrized by a variable phase. We present an analysis of the domain of control and support
our results by numerical calculations. 
\end{abstract}


\maketitle
\section{Introduction}
Since the last decade, the stabilization of unstable and chaotic systems has been a field of extensive research. A
variety of control schemes have been developed to control periodic orbits as well as steady states
\cite{OTT90,BAB02,SCH07}. A simple and efficient scheme, introduced by Pyragas \cite{PYR92}, is known as \textit{time
delay autosynchronization} (TDAS). This control method generates a feedback from the difference of the current state of
a system to its counterpart some time units $\tau$ in the past. Thus, the control scheme does not rely on a reference
system and has only a small number of control parameters, \textit{i.e.}, the feedback gain $K$ and time delay $\tau$. It
has been shown that TDAS can stabilize both unstable periodic orbits, e.g., embedded in a strange attractor
\cite{PYR92,BAL05}, and unstable steady states \cite{AHL04,ROS04a,HOE05}. In the first case, TDAS is most efficient if $\tau$
corresponds
to an integer multiple of the minimal period of the orbit. In the latter case, the method works best if the time delay
is related to an intrinsic characteristic timescale given by the imaginary part of the system's eigenvalue \cite{HOE05}.
A generalization of the original Pyragas scheme, suggested by Socolar et al. \cite{SOC94}, uses multiple time delays.
This \textit{extended time delay autosynchronization} (ETDAS) introduces a memory parameter $R$, which serves as a weight of states
further in the past. A variety of analytic results about time-delayed feedback control are also known
\cite{BLE96,JUS97,JUS98,PYR01}, for instance, in the case of long time delays \cite{YAN06} or the odd number limitation
\cite{NAK97}, which was refuted recently \cite{FIE07}.

Although there has been strong effort on the research on the original Pyragas method \cite{JUS03,JUS04}, much less is
known in the case of extended time-delayed feedback \cite{GAU98,PYR95a,PYR02,BEC02,UNK03,HOE03}. Recently it was shown that the
additional memory parameter introduces a second timescale which leads to a strong improvement of the stabilization
ability, for instance, arbitrary large correlations of stochastic oscillations without inducing a bifurcation
\cite{POM07}.

In the present paper, we apply the ETDAS control method, which was initially invented to stabilize periodic orbits, to
an unstable steady state realized as an unstable focus. This can be seen as a system close to, but above, a
supercritical Hopf bifurcation. As a modification, we consider also an additional control loop latency, a bandpass
filter, and different couplings.

This paper is organized as follows: In Sec.~\ref{sec:ETDAS}, we introduce our model equations and develop the
analytic tools used throughout the paper. Section~\ref{sec:phase} deals with a nondiagonal coupling implemented by a
rotational matrix. In Sec.~\ref{sec:latency}, we consider latency time effects, which arise if a time lag exists
between calculation of the control force and reinjection into the system. Section~\ref{sec:band-pass} introduces a
specific modification of ETDAS that includes a bandpass filtering of the control signal. This is important if high
frequency components are present in the system. Finally, we conclude with Sec.~\ref{sec:conclusion}.

\section{\label{sec:ETDAS}Extended time-delayed feedback}
We consider an unstable fixed point of focus type. Without loss of generality, the fixed point $z^*$ is located at the
origin. In complex center manifold coordinates $z$ the linearized system can be written as $\dot{z}(t)=(\lambda +
i\omega) z(t)$, where $\lambda$ and
$\omega$ are real numbers corresponding to the damping and oscillation frequency, respectively. The stability of the
steady state is determined by the sign of the real part of the complex eigenvalue $\lambda + i \omega$. Since we
consider an unstable focus, e.g., closely above a Hopf bifurcation, we choose the parameter $\lambda$ to be positive and
$\omega$ nonzero. Separated in real and imaginary parts, \textit{i.e.}, $z(t)=x(t)+iy(t)$, the dynamics of the system is
given by
\begin{eqnarray}
        \label{eq:dyn}
        \dot{\bf x}(t) = {\bf A} \; {\bf x}(t) - {\bf F}(t),
\end{eqnarray}
where $\bf x$ is the state vector composed of the real and imaginary part $x$ and $y$ of the variable $z$, the matrix
$\bf A$ denotes the dynamics of the uncontrolled system,
\begin{equation}
 \mathbf{A} =
\left( \begin{array}{cc}
\lambda & \omega \\
-\omega & \lambda
\end{array} \right) ,
\end{equation}
and $\bf F$ is
the ETDAS control force, which can be written in three equivalent forms
\begin{eqnarray}\label{eq:force}
	{\bf F}(t) &=&K \sum_{n=0}^{\infty}R^{n}\left[{\bf x}(t-n\tau)-{\bf x}\textbf{(}t-(n+1)\tau\textbf{)}\right] \\
	&=&K \left[{\bf x}(t)-(1-R)\sum_{n=1}^{\infty}R^{n-1}{\bf x}(t-n\tau)\right]\\
	&=&K \left[{\bf x}(t)-{\bf x}(t-\tau)\right]+R {\bf F}(t-\tau),\label{eq:force_recursive}
\end{eqnarray}
where $K$ and $\tau$ denote the (real) feedback gain and the time delay, respectively. $R \in (-1,1)$ is a memory
parameter that takes into account those states that are delayed by more than one time interval $\tau$. Note that $R=0$
yields the TDAS control scheme introduced by Pyragas \cite{PYR92}.

The control force applied to the $i$th component of the system consists only of contributions of the same component.
Thus, this realization is called diagonal coupling. We will consider a generalization to a nondiagonal coupling scheme
in Sec.~\ref{sec:phase}. The first form of the control force, Eq.~(\ref{eq:force}), indicates the noninvasiveness of the
ETDAS method because ${\bf x}^* (t-\tau)={\bf x}^*(t)$ if the fixed point is stabilized. The third form,
Eq.~(\ref{eq:force_recursive}), is suited best for an experimental implementation since it involves states further than
$\tau$ in the past only recursively.

While the stability of the fixed point in the absence of control is given by the eigenvalues of matrix ${\bf A}$,
\textit{i.e.}, $\lambda \pm i \omega$, one has to solve the following characteristic equation in the case of an ETDAS
control force:
\begin{eqnarray}
 \label{eq:char}
 \Lambda+K \frac{1-e^{-\Lambda\tau}}{1-R e^{-\Lambda\tau}} = \lambda \pm i\omega.
\end{eqnarray}
Due to the presence of the time delay $\tau$, this characteristic equation becomes transcendental and possesses an
infinite but countable set of complex solutions $\Lambda$. In the case of TDAS, \textit{i.e.}, $R=0$, the
characteristic equation can be solved analytically in terms of the Lambert function \cite{WRI49,HAL71,HOE05,AMA05}. We
stress that for nonzero memory parameter $R$, however, such a compact analytic expression is not possible. Thus, one
has to solve Eq.~(\ref{eq:char}) numerically.

\begin{figure}[ht]
\includegraphics[width= \linewidth,angle=0]{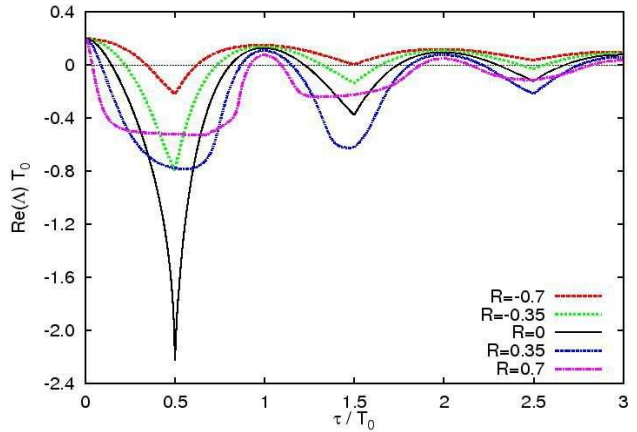}
\caption{\label{fig:ReLambda}(Color online) Largest real part of the complex eigenvalues $\Lambda$ as a function of $\tau$ for
different values of $R$. The red, green, black, blue, and magenta curves correspond to $R= -0.7, -0.35,0,0.35,0.7$,
respectively. The parameters of the unstable focus are chosen as $\lambda=0.1$ and $\omega=\pi$ which yields an
intrinsic period $T_0= 2 \pi /\omega=2$. The feedback gain $K$ is fixed at $KT_0=0.6$.}
\end{figure}

Figure~\ref{fig:ReLambda} depicts the dependence of the largest real parts of the eigenvalue $\Lambda$ upon the time
delay $\tau$ according to Eq.~(\ref{eq:char}) for different memory parameters $R$ and fixed feedback gain $K=0.3$. The
red, green, black, blue, and magenta curves of $\Re(\Lambda)$ correspond to $R= -0.7, -0.35,0,0.35,0.7$, respectively.
The parameters of the unstable focus are chosen as $\lambda = 0.1$ and $\omega=\pi$. Note that the time delay $\tau$ is
given in units of the intrinsic period $T_0= 2 \pi /\omega$. When no control is applied to the system, \textit{i.e.},
$\tau = 0$, all curves start at $\lambda$ which corresponds to the real part of the uncontrolled eigenvalue. For
increasing time delay, the real part of $\Lambda$ decreases and eventually changes sign. Thus, the fixed point becomes
stable. Note that there is a minimum of $\Re(\Lambda)$ indicating strongest stability if the time delay $\tau$ is equal
to half the intrinsic period. For larger values of $\tau$, the real part increases and becomes positive again. Hence,
the system loses its stability. Above $\tau=T_0$, the cycle is repeated but the minimum of $\Re(\Lambda)$ is not so
deep. The control method is less effective because the system has already evolved further away from the fixed
point. For vanishing memory parameter $R=0$ (TDAS), the minimum is deepest, however, the control interval,
\textit{i.e.}, values of $\tau$ with negative real parts of $\Lambda$, increases for larger $R$. Therefore the ETDAS
control method is superior in comparison to the Pyragas scheme.

\begin{figure}[ht]
\includegraphics[width= \linewidth,angle=0]{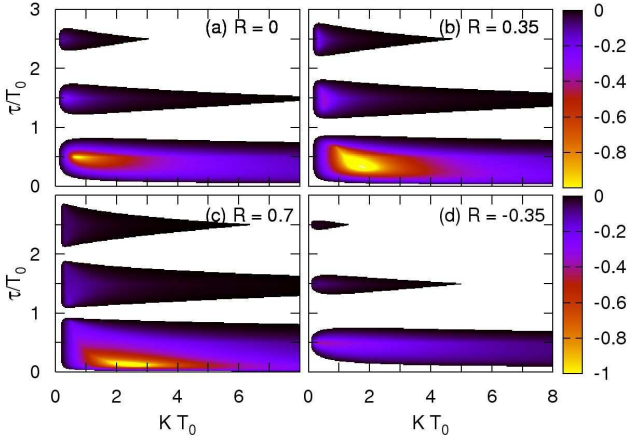}
\caption{\label{fig:Ktau}(Color online) Domain of control in the $(K,\tau)$ plane for different values of $R: 0, 0.35,
0.7, -0.35$ in panels (a), (b), (c), and (d), respectively. The greyscale (color online) shows only negative values
of the largest real part of the complex eigenvalues $\Lambda$ according to Eq.~(\ref{eq:char}). The parameters of the
system are as in Fig.~\ref{fig:ReLambda}.}
\end{figure}

Figure~\ref{fig:Ktau} shows the domain of control in the plane parametrized by the feedback gain $K$ and time delay
$\tau$ for different values of $R: 0, 0.35, 0.7, -0.35$ in panels (a), (b), (c), and (d), respectively. The color code
indicates only negative values of the largest real parts of the complex eigenvalue $\Lambda$. Therefore,
Fig.~\ref{fig:ReLambda} can be understood as a vertical cut through Fig.~\ref{fig:Ktau} for a fixed value of $KT_0
=0.6$. Each panel displays several islands of stability which shrink for larger time delays $\tau$. Note that no
stabilization is possible if $\tau$ is equal to an integer multiple of the intrinsic period $T_0$. The domains of
control become larger if the memory parameter $R$ is closer to $1$.

In order to obtain some analytic information of the domain of control, it is helpful to separate the characteristic
equation~(\ref{eq:char}) into real and imaginary parts. This yields using $\Lambda=p+iq$:
\begin{eqnarray}
	\label{eq:real_and_imag}
	K (1-e^{-p\tau}\cos{q\tau}) & = & \lambda - p - R e^{-p\tau}\nonumber \\
	\times [ (\lambda &-& p) \cos{q\tau} \pm (\omega-q) \sin{q\tau}]
\end{eqnarray} 
and
\begin{eqnarray}
	\label{eq:real_and_imag2}
	K e^{-p\tau} \sin{q\tau} & = & \pm (\omega-q) + R e^{-p\tau} \nonumber \\
	\times [(\lambda &-& p) \sin{q\tau} \pm (\omega-q) \cos{q\tau}].
\end{eqnarray}
The boundary of the domain of controls is determined by a vanishing real part of $\Lambda$, \textit{i.e.}, $p=0$. With
this constraint, Eqs.~(\ref{eq:real_and_imag}) and (\ref{eq:real_and_imag2}) can be rewritten as
\begin{eqnarray}
	\label{eq:real_and_imag_p=0}
	K (1-\cos{q\tau}) & = & \lambda - R [ \lambda \cos{q\tau} \pm (\omega-q) \sin{q\tau}],\\
	K \sin{q\tau} & = & \pm (\omega-q) + R [\lambda \sin{q\tau} \pm (\omega-q) \cos{q\tau}] \nonumber.
\end{eqnarray}

At the threshold of control ($p =0, q=\omega$), there is a certain value of the time delay, which will serve as a reference in the following Section, given by
\begin{equation}
	\label{eq:opt_tau}
	\tau=\frac{\left(2 n +1\right) \pi }{\omega} = \left(n+\frac{1}{2}\right)T_0
\end{equation}
where $n$ is any nonnegative integer. For this special choice of the time delay, the range of possible feedback gains $K$ in the domain of control becomes largest as can be seen in Fig.~\ref{fig:Ktau}. Hence, we will refer to this $\tau$ value as optimal time delay in the following. The minimum feedback gain at this $\tau$ can be obtained:
\begin{equation}
 	\label{eq:Kmin}
	K_{min} (R) =\frac{\lambda \left(1+R\right)}{2}.
\end{equation}

Extracting an expression for $\sin(q\tau)$ from Eq.~(\ref{eq:real_and_imag_p=0}) and inserting it into the equation for the imaginary part leads after some algebraic manipulation to a general dependence of $K$ on the imaginary part $q$ of $\Lambda$
\begin{eqnarray}\label{eq:Kq}
	K(q) =\frac{(1+R)\left[\lambda^2+(\omega-q)^2\right]}{2 \lambda}.
\end{eqnarray} 
Taking into account the multivalued properties of the arcsine function, this yields in turn analytical expressions of the time delay in dependence on $q$
\begin{eqnarray}
	\tau_1(q) &=& \frac{\arcsin\left(\frac{2\lambda(1-R^2)(\omega-q)}{\lambda^2(1-R^2)^2+(\omega-q)^2(1+R)^2}\right)+2n\pi}{q},\\
	\tau_2(q) &=& \frac{-\arcsin\left(\frac{2\lambda(1-R^2)(\omega-q)}{\lambda^2(1-R^2)^2+(\omega-q)^2(1+R)^2}\right)+(2n+1)\pi }{q}, \nonumber
\end{eqnarray} 
where $n$ is a nonnegative integer. Together with Eq.~(\ref{eq:Kq}), these formulas describe the boundary of the domain of control in Fig.~\ref{fig:Ktau}. Note that two expressions $\tau_1$ and $\tau_2$ are necessary to capture the complete boundary. The case of TDAS control was analyzed in \cite{YAN06} and is included as special choice of $R=0$.

\begin{figure}[ht]
\includegraphics[width= \linewidth,angle=0]{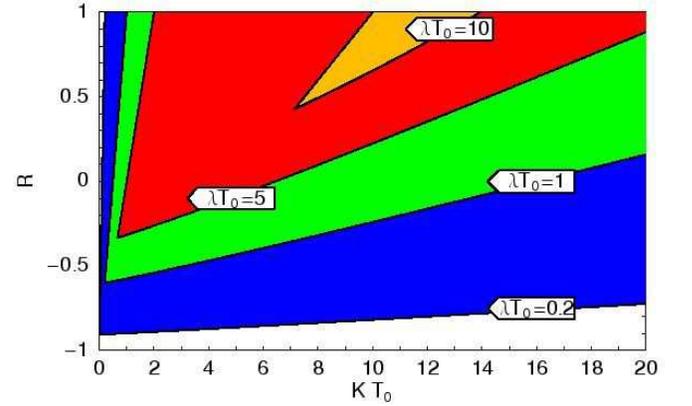}
\caption{\label{fig:RK}(Color online) Domain of control in the $(K,R)$ plane for different values of $\lambda$. The
blue, red, green, and yellow domains correspond to $\lambda T_0=0.2,1,5,$ and $10$, respectively, as indicated. The time
delay is
chosen as $\tau=T_0/2$ and $\omega=\pi$.}
\end{figure}

For a better understanding of effects due to the memory parameter $R$, it is instructive to consider the domain of
control in the plane parametrized by $R$ and the feedback gain $K$. The results can be seen in Fig.~\ref{fig:RK}, where
the blue, red, green, and yellow areas correspond to the domain of control for $\lambda T_0=0.2,1,5,$ and $10$,
respectively. The other system parameter is chosen as $\omega = \pi$. We keep the time delay constant at $\tau =
T_0/2$. Note that the $K$ interval for successful control increases for larger values of $R$. In fact, while the
original Pyragas scheme, \textit{i.e.}, $R=0$, fails for $\lambda T_0 =10$, the ETDAS method is still able to stabilize
the fixed point.
The upper left boundary corresponds to Eq.~(\ref{eq:Kmin}). The lower right boundary can be described by a
parametric representation which can be derived from the characteristic equation~(\ref{eq:char}):
\begin{eqnarray}
	\label{eq:KRparametric_1}
	R & = & \frac{\lambda\tau - \vartheta\tan{(\vartheta/2)}}{\lambda\tau +
\vartheta\tan{(\vartheta/2)}}, \\
	\label{eq:KRparametric_2}
	K \tau & = & \frac{\vartheta^{2} + \left(\lambda\tau\right)^{2}}{\lambda\tau +
\vartheta\tan{(\vartheta/2)}}, 
\end{eqnarray}
where we used the abbreviation $\vartheta=\left(q-\omega\right)\tau$ for notational convenience. The range of $\vartheta$ is given by $\vartheta \in \left[ 0,\pi \right) $. A linear approximation  leads to an analytic dependence of $R$ an the feedback gain $K$ given by a function $R(K)$ instead of the parametric
equations~(\ref{eq:KRparametric_1}) and (\ref{eq:KRparametric_2}). A Taylor expansion around $\vartheta=\pi$ yields
\begin{equation}
	\label{eq:KRlinear}
	K_{max} (R) = \frac{\lambda^{2}+\pi^{2}}{2\lambda} \left(R+1\right)+2 \left(R-1\right).
\end{equation}

\begin{figure}[ht]
\includegraphics[width= \linewidth,angle=0]{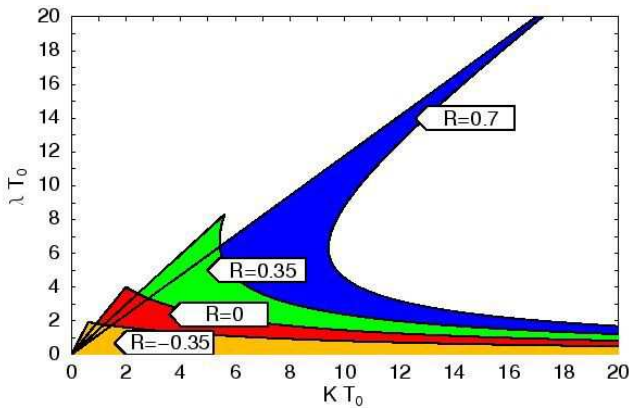}
\caption{\label{fig:lambdaK}(Color online) Domain of control in the $(K,\lambda)$ plane for different memory parameters
 $R$. The yellow, red, green, and blue areas correspond to $R=-0.35$, $0$ (TDAS), $0.35$, and $0.7$, respectively. The
time delay is fixed at $\tau = T_0/2$.}
\end{figure}
Another representation of the superior control ability of ETDAS is depicted in Fig.~\ref{fig:lambdaK}. The domain of
control is given in the $(K,\lambda)$ plan for different values of $R$. The yellow, red, green, and blue areas refer to
$R=-0.35$, $0$ (TDAS), $0.35$, and $0.7$, respectively. The time delay is chosen as $\tau=T_0/2$. One can see that for
increasing $R$, the ETDAS method can stabilize systems in a larger $\lambda$ range. However, the corresponding $K$
interval for successful control can become small. See, for instance, the blue area $(R=0.7)$ for large $\lambda$. A
similar behavior was found in the case of stabilization of an unstable periodic orbit by ETDAS \cite{JUS99}. We stress that, as in the
case of periodic orbits, the boundaries of the shaded areas can be calculated analytically from the following expression:
\begin{eqnarray}\label{eq:Klambda}
	K \tau &=& \frac{(1-R) \vartheta}{\tan(\vartheta/2)} \left[\left(\frac{1+R}{1-R}\right)^2+\tan^2(\vartheta/2)\right],\\
	\lambda \tau &=& \frac{\vartheta}{\tan(\vartheta/2)} \, \left(\frac{1+R}{1-R}\right),
\end{eqnarray}
where we used $\vartheta=\left(q-\omega\right)\tau$ with $\vartheta \in [0,\pi)$ as in Eqs.~(\ref{eq:KRparametric_1}) and
 (\ref{eq:KRparametric_2}). The maximum value for $\lambda$, which can be stabilized, is given by the special case $\vartheta=0$:
\begin{eqnarray}
	\lambda_{max} \tau = 2  \, \frac{1+R}{1-R}.
\end{eqnarray} 

In this Section, we have focussed our discussion to the ETDAS method in the simplest realization of diagonal coupling.
In the following, we will investigate the effects of nondiagonal coupling introduced by a variable phase.

\section{\label{sec:phase}Phase dependent coupling}
Time-delayed feedback has been widely used in optical systems both to study the intrinsic dynamics and to control the
stabilization of, for instance, a laser device \cite{BIE93,PIE94,USH04,BAU04,TRO06,GAV06}. In these systems, the optical
phase is an additional degree of freedom. We consider this additional control parameter as a generalization of the ETDAS
feedback scheme using a nondiagonal coupling as opposed to the diagonal coupling discussed in the previous Section. This
nondiagonal coupling is realized by introducing a coupling matrix containing a variable phase $\varphi$:
\begin{eqnarray}
 \label{eq:uss_phase}
 \left(
 \begin{array}{c}
  \dot{x} \\
  \dot{y}
 \end{array}
 \right) &=& \left(
 \begin{array}{cc}
  \lambda &\omega \\
  -\omega & \lambda
 \end{array}
 \right) \left(
 \begin{array}{c}
  x \\
  y
 \end{array}
 \right)  \nonumber\\
 &&- 
 \left(
 \begin{array}{cc}
  \cos\varphi & - \sin\varphi\\
  \sin\varphi & \cos\varphi
 \end{array}
 \right)
 {\bf F}(t),
\end{eqnarray}

In optical systems like semiconductor lasers with external optical feedback \cite{FIS00a,SCH06a}, this feedback phase
can be seen as the phase of the electric field. Experimentally, this phase of the feedback can be varied by tuning an
external Fabry-Perot cavity. It has also been demonstrated that a feedback phase plays an important role in the
suppression of collective synchrony in a globally coupled oscillator network \cite{ROS04}.

\begin{figure}[ht]
\includegraphics[width= \linewidth,angle=0]{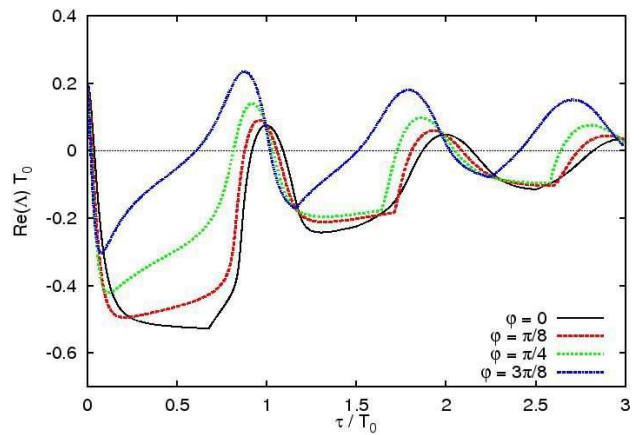}
\caption{\label{fig:ReLambda_phase}(Color online) Largest real part of the eigenvalues $\Lambda$ as a function of $\tau$ for
different phases $\varphi$. The black, red, green, and blue curves correspond to $\varphi=0$, $\pi / 8$, $\pi / 4$, and $3 \pi /8$, respectively. The other control parameters are fixed as $R=0.7$ and $KT_0=0.6$. The parameters of the system are as in Fig.~\ref{fig:ReLambda}.}
\end{figure}

\begin{figure}[ht]
\includegraphics[width= \linewidth,angle=0]{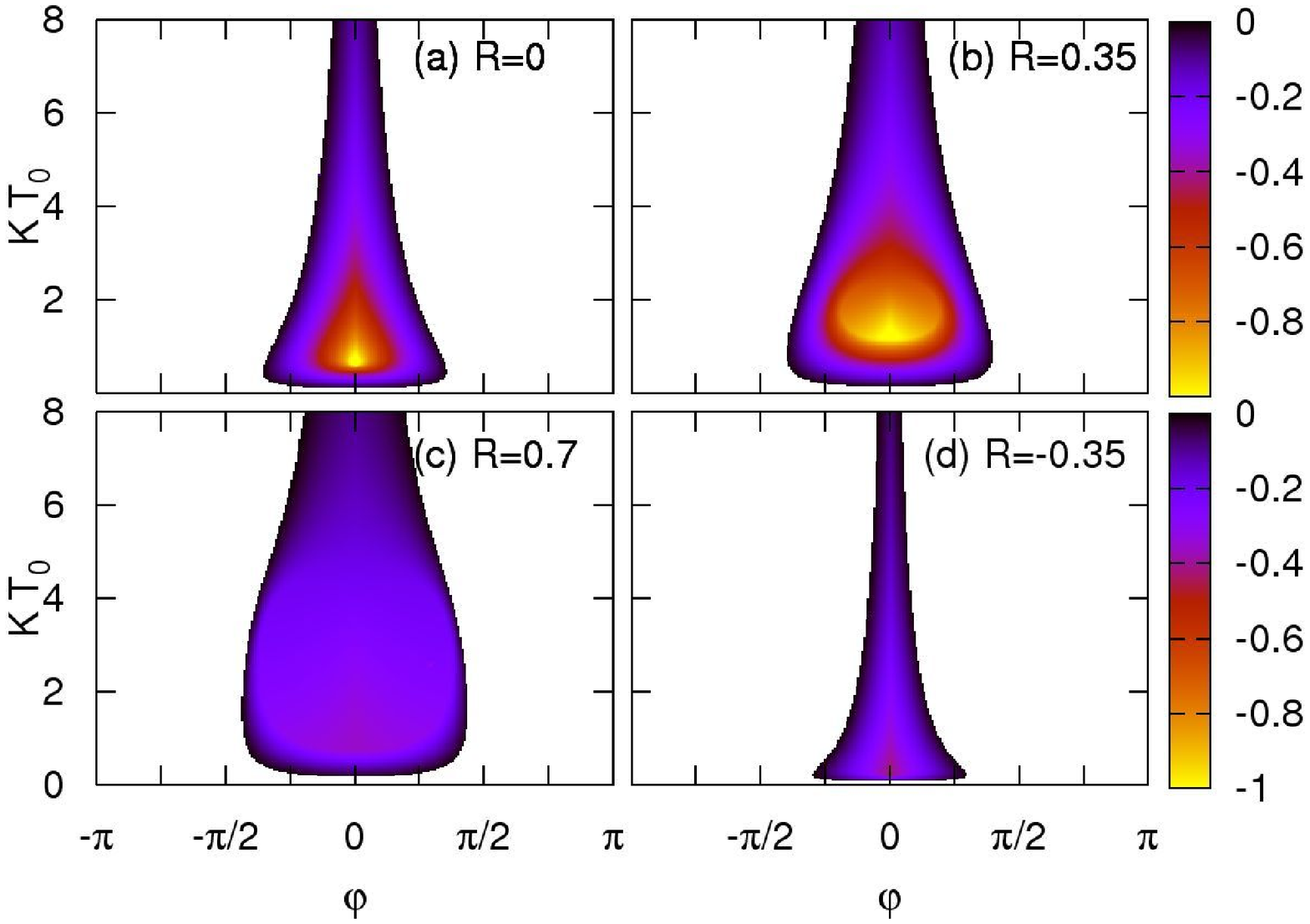}
 \caption{\label{fig:Kphi_opt}(Color online) Domain of control in the $(\varphi,K)$ plane for optimal time delay $\tau =
T_0/2$. Panels (a), (b), (c), and (d) correspond to a memory parameter $R$ of $0$, $0.35$, $0.7$, and $-0.35$,
respectively. The color code shows the largest real part of the complex eigenvalues $\Lambda$ as given by
Eq.~(\ref{eq:char_phase}). The parameters of the system are as in Fig.~\ref{fig:ReLambda}.}
\end{figure}

\begin{figure}[ht]
\includegraphics[width= \linewidth,angle=0]{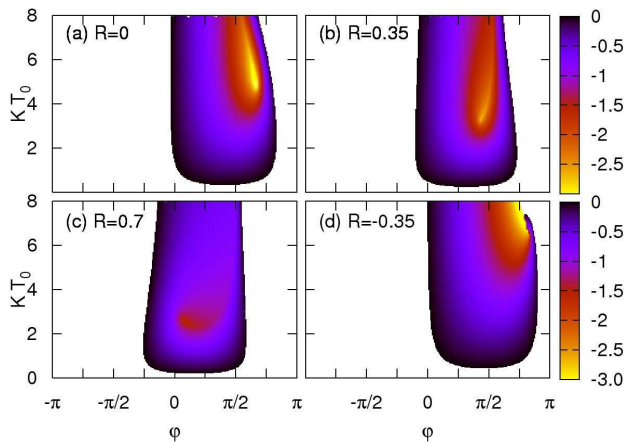}
 \caption{\label{fig:Kphi_bad}(Color online) Domain of control in the $(\varphi,K)$ plane for time delay $\tau = 0.1
T_0$. Panels (a), (b), (c), and (d) correspond to a memory parameter $R$ of $0$, $0.35$, $0.7$, and $-0.35$,
respectively. The color code shows the largest real part of the complex eigenvalues $\Lambda$ as given by
Eq.~(\ref{eq:char_phase}). Only negative values are depicted. The parameters of the system are as in
Fig.~\ref{fig:ReLambda}.}
\end{figure}

The stability of the fixed point is again given by the largest real part of the complex eigenvalues $\Lambda$, which are calculated as the solutions of the following modified characteristic equation:
\begin{eqnarray}
 \label{eq:char_phase}
 \Lambda+K e^{\mp i\varphi} \frac{1-e^{-\Lambda\tau}}{1-R e^{-\Lambda\tau}} = \lambda \pm i\omega.
\end{eqnarray}
Note that this equation differs from the characteristic equation in the diagonal case (see Sec.~\ref{sec:ETDAS}) by an 
additional exponential term. This is due to the choice of phase dependent coupling by a rotational matrix. We stress that 
a similar phase factor has recently been used \cite{FIE07} to overcome some topological limitation of time-delayed feedback control 
known as \textit{odd number limitation theorem}, which refers to the case of an unstable periodic orbit with an odd number 
of real Floquet multipliers larger than unity. \cite{NAK97,NAK98}

In analogy to Sec. \ref{sec:ETDAS} a minimum feedback gain can be calculated:
\begin{equation}
K_{min} = \frac{\lambda\left(1+R\right)}{2\cos{\varphi}} .
\label{eqn:phase_k_min}
\end{equation}
Note that the time delay that corresponds to this value of $K_{min}$ is no longer given by Eq.~(\ref{eq:opt_tau}) and is not the optimal time delay in the general case of nonzero phase. Nevertheless, Eq.~(\ref{eqn:phase_k_min}) can be used as a coarse estimate of the minimum feedback gain for the regime of small values of $\varphi$, if the time delay is chosen as Eq.~(\ref{eq:opt_tau}).

\begin{figure}[ht]
\includegraphics[width= \linewidth,angle=0]{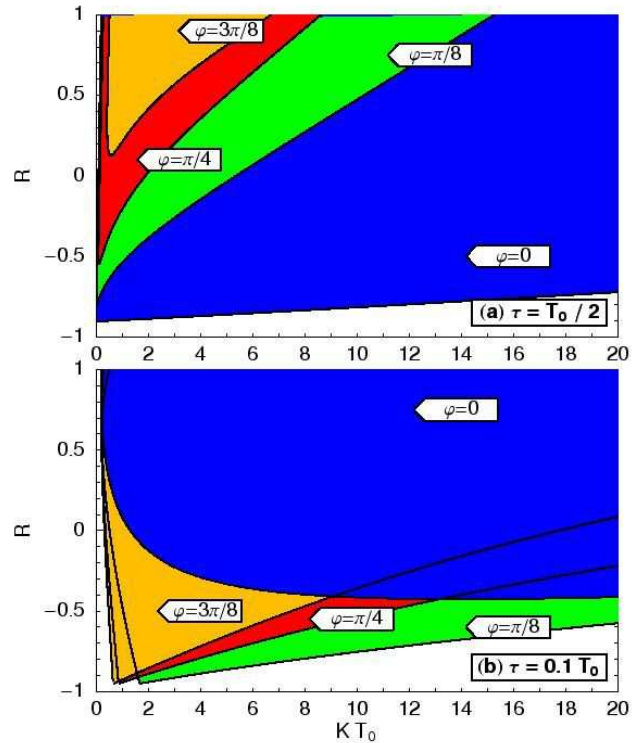}
\caption{\label{fig:KR_phase}(Color online) Domain of control in the $(K,R)$ plane for different values
of the feedback phase $\varphi$. The blue, green, red, and yellow areas correspond to $\varphi=0$, $\pi/8$, $\pi/4$, and
$3\pi/8$, respectively. Panel (a) displays the domain of control for optimal $\tau=T_0/2$; panel (b) for $\tau=0.1 T_0$.
The parameters of the system are as in Fig.~\ref{fig:ReLambda}.}
\end{figure}

Figure~\ref{fig:ReLambda_phase} depicts the dependence of the largest real part of the eigenvalues $\Lambda$ on the time
delay $\tau$ for fixed values of $R=0.7$ and $KT_0=0.6$, but different values of the phase. The black, red, green, and
blue curves correspond to $\varphi=0$, $\pi/8$, $\pi/4$, and $3\pi/8$, respectively. It can be observed that the control
is overall less effective for larger $\varphi$, as the curves are shifted up towards positive reals parts for increasing
the phase. The range of possible values for the time delay shrinks. The optimal time delay is shifted towards smaller
values for larger $\varphi$, which can be seen for the case of $\varphi=3\pi/8$, where the optimal time delay is in the
range of $\tau=0.1 T_{0}$ instead of $0.5 T_{0}$, which was the optimal time delay for $\varphi=0$ according to
Eq.~(\ref{eq:opt_tau}).

The four-dimensional parameter space is now given by the feedback gain $K$, time delay $\tau$, memory parameter $R$, and feedback phase $\varphi$. At first, we consider the domain of control in the plane parametrized by $K$ and $\varphi$. Hence, we keep the other remaining control parameters $R$ and $\varphi$ fixed. Figures~\ref{fig:Kphi_opt} and \ref{fig:Kphi_bad} show the domain of control for a time delay  of $T_0/2$ and $0.1 T_0$, respectively. In each figure, the memory parameter $R$ is chosen as $R=0$, $0.35$, $0.7$, and $-0.35$ in panels (a), (b), (c), and (d), respectively. The color code corresponds to the largest real part of the complex eigenvalues as calculated from Eq.~(\ref{eq:char_phase}). Only negative values are depicted, \textit{i.e.}, those combinations of $K$ and $\varphi$ for which the control scheme is successful. Note that the case $R=0$ corresponds to the TDAS control method \cite{SCH06a}. An increase of the memory parameter $R$ leads to a larger domain of control. Even though the system can be stabilized for a larger range of $K$ and $\varphi$, the system becomes over all less stable, since the real part of $\Lambda$ is closer to zero. For negative values of $R$, the domain of control shrinks. Note that also in the case of non-optimal time delay as in Fig.~\ref{fig:Kphi_bad} the range of choice for possible feedback gain and phase is enlarged.

For a better understanding of the effects of the feedback phase, Fig.~\ref{fig:KR_phase} depicts the domain of control
in the $(K,R)$ plane. The blue, green, red, and yellow areas correspond to successful control for $\varphi=0$, $\pi/8$,
$\pi/4$, and $3\pi/8$, respectively. Panel (a) shows the case of optimal time delay, \textit{i.e.}, $\tau=T_0/2$; panel
(b) displays the case of $\tau=0.1 T_0$. Note that an increase of $\varphi$ leads to a smaller domain of control in the
case of $\tau=T_0/2$. This effect, however, is reversed for non-optimal choices of $\tau$, where the phase $\varphi$
compensates for the bad choice of the time delay. Thus, control is possible again, for instance, in the TDAS case
($R=0$). Following the strategy introduced in Sec.~\ref{sec:ETDAS}, one can derive also in the case $\varphi\neq 0$
parametric formulas for the the boundary of the domain of control:
\begin{widetext}
\begin{eqnarray}
	R &=& \frac{\vartheta\left[\cos{\left(\omega\tau+\vartheta+\varphi\right)}-\cos{\varphi}\right]+\lambda\tau\left[\sin{\varphi} -\sin{\left(\omega\tau+\vartheta+\varphi\right)}\right]}{\vartheta\left[\cos{\varphi} -\cos{\left(\omega\tau+\vartheta+\varphi\right)}\right]-\lambda\tau\left[\sin{\varphi}+\sin{\left(\omega\tau+\vartheta+\varphi\right)}\right]} ,
	\\
	K\tau &=& \frac{\left(\vartheta^{2}+\lambda^{2}\tau^{2}\right)\cos{\left[\frac{1}{2}\left(\omega\tau+\vartheta\right)\right]}}{\lambda\tau\cos
{\left[\frac{1}{2}\left(\omega\tau+\vartheta\right)-\varphi\right]}-\vartheta\sin{\left[\frac{1}{2}\left(\omega\tau+\vartheta\right)-\varphi\right]}} 
\end{eqnarray} 
\end{widetext}

We stress that it is possible to derive expression of $K(q)$ and $\tau(q)$ similar to Eqs.~(\ref{eq:Kq}) and (\ref{eq:Klambda}) also in the case of $\varphi \neq 0$. These calculations are lengthy and do not produce more insight and thus are omitted here.

\section{\label{sec:latency}Control loop latency}
After the investigation of nondiagonal coupling, we consider in this Section an additional control loop latency. This latency is associated with time required for the generation of the control force and its reinjection into the system. In an optical realization, for instance, the latency time is given by the propagation time of the light between the laser and the Fabry-Perot control device. We stress that in the case of unstable periodic orbits, Just has shown that longer latency times reduce the control abilities of the time-delayed feedback of TDAS type \cite{JUS99}. Similar results were found for ETDAS \cite{HOE03}.

\begin{figure}[ht]
\includegraphics[width= \linewidth,angle=0]{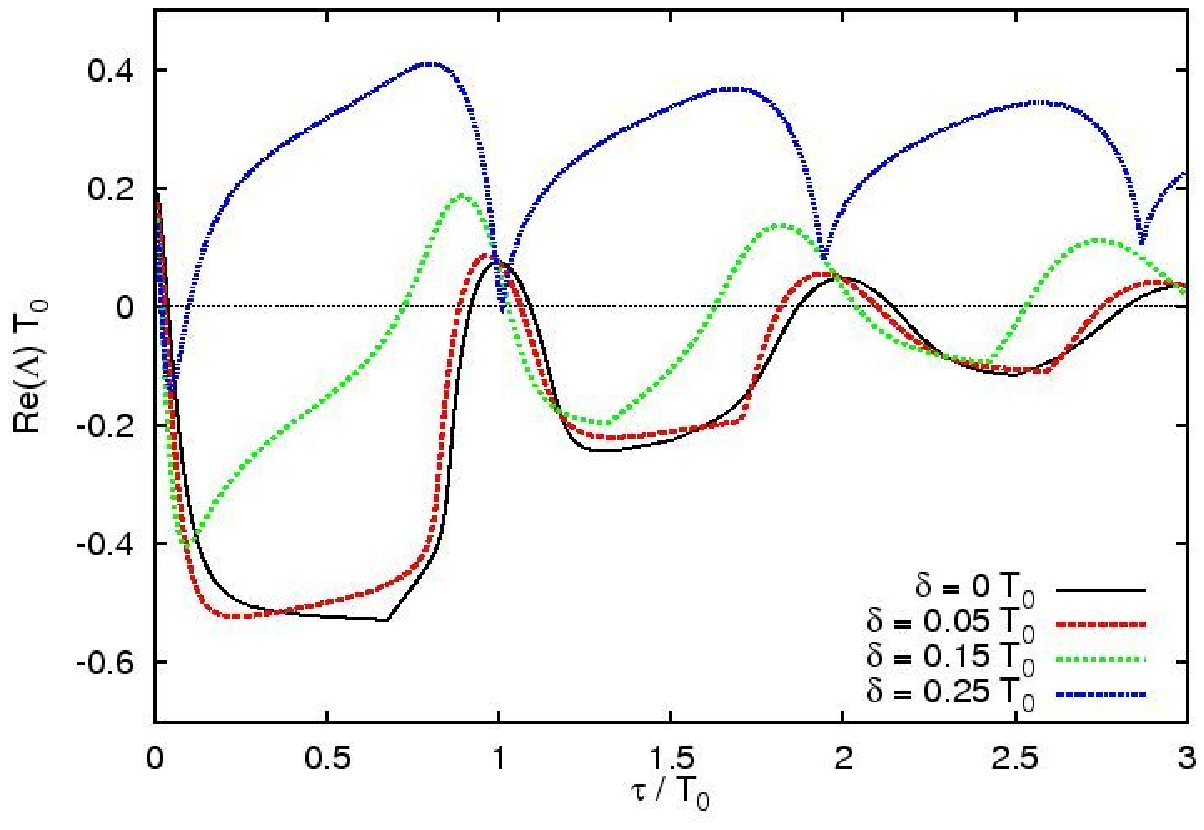}
\caption{\label{fig:ReLambda_latency}(Color online) Largest real part of the eigenvalues $\Lambda$ as a function of $\tau$ for
different latency times $\delta$. The black, red, green, and blue curves correspond to $\delta=0$, $0.05 T_{0}$, $0.15 T_{0}$, and $0.25 T_{0}$, respectively. The other control parameters are fixed as $R=0.7$ and $KT_0=0.6$. The parameters of the system are as in Fig.~\ref{fig:ReLambda}.}
\end{figure}

The latency time $\delta$ acts as an additional time delay in all arguments of the control force of Eq.~(\ref{eq:force}). Using the recursive form of ${\bf F}$ as given by Eq.~(\ref{eq:force_recursive}), this yields 
\begin{eqnarray}
	\label{eq:uss_latency}
        {\bf F}(t) 
	&=& \left[{\bf x}(t-\delta)-{\bf x}(t-\delta-\tau)\right]+R {\bf F}(t-\tau).
\end{eqnarray}
The characteristic equation (\ref{eq:char}) is now modified by an additional exponential factor
\begin{eqnarray}
	\label{eq:char_latency} 
	\lambda \pm i \omega &=& \Lambda + K e^{- \Lambda \delta} \frac{1-e^{-\Lambda\tau}}{1-R e^{-\Lambda\tau}}.
\end{eqnarray}
In contrast to the previous Section, this exponential term depends on the eigenvalue $\Lambda$ itself.

Figure~\ref{fig:ReLambda_latency} depicts the dependence of the largest real part of the eigenvalues $\Lambda$ on the
time delay $\tau$ for fixed values of $R=0.7$ and $KT_0=0.6$, but different latency times. The black, red, green, and
blue curves correspond to $\delta=0$, $0.1$, $0.3$, and $0.5$, respectively. It can be seen that the control scheme is
less successful for longer latency times. The $\tau$ interval with negative real parts of $\Lambda$ becomes smaller.  In
the case of $\delta=0.25 T_0$, for instance, control can only be achieved in a narrow range of small $\tau$ and the
second minimum does not reach down to negative $\Re(\Lambda)$ anymore. In addition, the minima of the real parts are
distorted and shifted towards smaller time delays.

\begin{figure}[ht]
\includegraphics[width= \linewidth,angle=0]{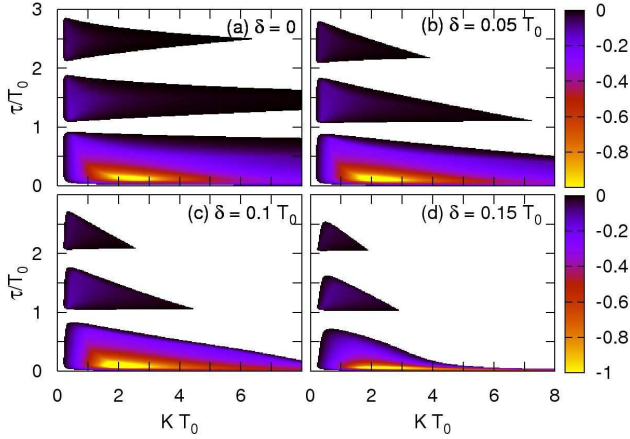}
\caption{\label{fig:Ktau_latency}(Color online) Domain of control in the $(\tau,K)$  plane for different values of
the latency time $\delta$ and fixed $R=0.7$. Panels (a), (b), (c), and (d) correspond to values of $\delta=0$, $0.05 T_{0}$, $0.1 T_{0}$, and $0.15 T_{0}$, respectively, where the time delay is fixed at $\tau = T_0/2$. The parameters of the system are as in Fig.~\ref{fig:ReLambda}.}
\end{figure}

Taking also a varying feedback gain $K$ into account, the domain of control can be seen in Fig.\ref{fig:Ktau_latency}.
The other control parameters are fixed at $\tau=T_0/2$ and $R=0.7$. Panels (a), (b), (c), and (d) correspond to values
of $\delta=0$, $0.1 \tau$, $0.2 \tau$, and $0.3 \tau$, respectively. As in Figs.~\ref{fig:Kphi_opt} and
\ref{fig:Kphi_bad} of the previous Section, the color code corresponds to the largest real part of the complex
eigenvalues which are calculated from Eq.~(\ref{eq:uss_latency}). Note that only negative values are depicted. For
increasing latency time, the domains of control shrink. Similar to the discussion in the one-dimensional projection of
Fig.~\ref{fig:ReLambda_latency}, the islands are distorted towards smaller time delays.

Separating the characteristic equation (\ref{eq:char_latency}) into real and imaginary part, one can derive, in analogy to Sec.~\ref{sec:ETDAS}, an expression for the minimum feedback gain 
\begin{eqnarray}
	K_{min}(\delta) &=& \frac{\lambda (1+R)}{\cos[(2n+1)\pi \delta/\tau]},
\end{eqnarray} 
which is consistent with the TDAS case investigated in \cite{HOE05}.

\begin{figure}[ht]
\includegraphics[width= \linewidth,angle=0]{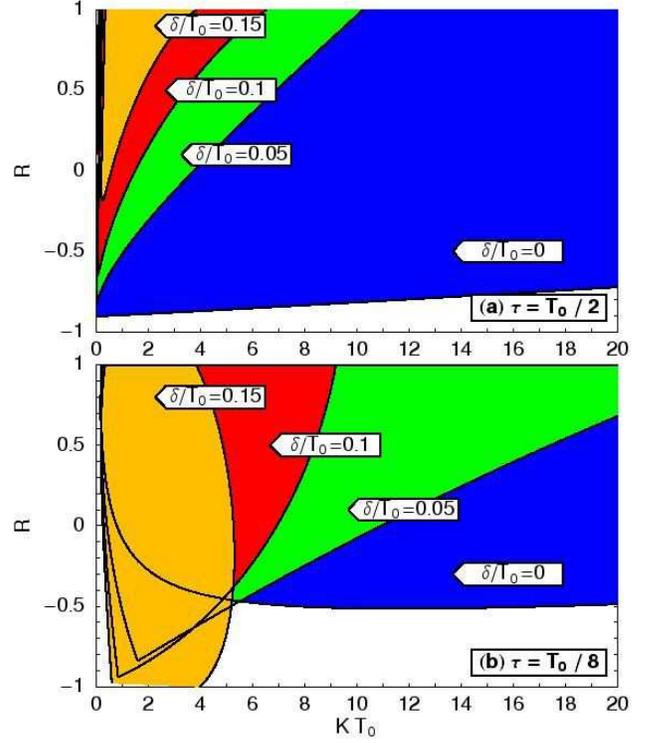}
\caption{\label{fig:KR_latency}(Color online) Domain of control in the $(K,R)$ plane for different values of the latency time $\delta$. The blue, green, red, and yellow areas refer to values of $\delta= 0$, $0.05 T_0$, $0.1T_0$, and $0.15 T_0$, respectively. Panel (a) corresponds to an optimal time delay $\tau = T_0 /2$, panel (b) to $\tau=T_0/8$. The parameters of the system are as in Fig.~\ref{fig:ReLambda}.}
\end{figure}

As another two-dimensional projection of the parameter space, Fig.~\ref{fig:KR_latency} displays the domain of control in the  $(K,R)$ plane for different values of the latency time $\delta$. The blue, green, red, and yellow areas refer to values of $\delta= 0$, $0.05 T_0$, $0.1T_0$, and $0.15 T_0$, respectively. Similar to Fig.~\ref{fig:KR_phase}, panel (a) shows the case of optimal choice of the time delay $\tau=T_0/2$ and panel (b) refers to $\tau=T_0/8$. In the first case, the domain of control shrinks considerably for increasing $\delta$, whereas in the latter case, this change is less pronounced.

Similar to the previous Sections, it is possible to derive a parametric expression for the boundary of the domain of control [$R(\vartheta)$ and $K(\vartheta)$]:
\begin{widetext}
\begin{eqnarray}
R &= & \frac{-\theta\cos{\left[\left(\omega\tau+\theta\right)\frac{\delta}{\tau}\right]} + \theta \cos{\left[\left(\omega\tau+\theta\right)\left(1+\frac{\delta}{\tau}\right)\right]} + \lambda\tau \left( \sin{\left[\left(\omega\tau+\theta\right)\frac{\delta}{\tau}\right]} -  \sin{\left[\left(\omega\tau+\theta\right)\left(1+\frac{\delta}{\tau}\right)\right]} \right) }{\theta\cos{\left[\left(\omega\tau+\theta\right)\frac{\delta}{\tau}\right]} - \theta \cos{\left[\left(\omega\tau+\theta\right)\left(1-\frac{\delta}{\tau}\right)\right]} - \lambda\tau \left( \sin{\left[\left(\omega\tau+\theta\right)\frac{\delta}{\tau}\right]} +  \sin{\left[\left(\omega\tau+\theta\right)\left(1-\frac{\delta}{\tau}\right)\right]} \right)} ,\\
K \tau & = & \frac{\left[\theta^{2} + \left(\lambda\tau\right)^{2}\right] \cos{\left[\frac{1}{2}\left(\omega\tau+\theta\right)\right)}}{ \lambda\tau\cos{\left[\left(\frac{\delta}{\tau}-\frac{1}{2}\right)\left(\omega\tau+\theta\right)\right]} + \theta\sin{\left[\left(\frac{\delta}{\tau}-\frac{1}{2}\right)\left(\omega\tau+\theta\right)\right]}} .
\label{eqn:latency_k_r}
\end{eqnarray}
\end{widetext}

\section{\label{sec:band-pass}Bandpass filtering}
In the context of semiconductor lasers, it has been shown both theoretically and experimentally that delayed, bandpass
filtered optical feedback is able to suppress the undamping of relaxation oscillations \cite{YOU99,FIS00a}.
Recently, filtering of an optical feedback signal \cite{ERZ06} was also investigated for laser models of Lang-Kobayashi type
\cite{LAN80b}. In addition, filtered feedback has also been proven important in the investigation of a Hopf
bifurcation \cite{ILL05}, and in the stabilization of unstable periodic orbits by ETDAS in semiconductor superlattices
\cite{SCH03a}. In order to model this type of modification of the control force, a bandpass filter acting on
the feedback force is introduced as a Lorentzian in the frequency domain with the transfer function
\begin{equation}
 T(\omega) = \frac{1}{1+i\frac{\omega-\omega_{0}}{\gamma}} ,
 \label{eqn:bandpass_T_omega}
\end{equation}
where $\omega_{0}$ denotes the peak of the transfer function and $\gamma$ is half the full width at half maximum
of the function. To introduce the filter into the system in the time domain, one can add two additional differential
equations to Eq.(\ref{eq:dyn}) such
that the
original two-dimensional system becomes four-dimensional:
\begin{eqnarray}
\label{eqn:filtered_system}
 \dot{\mathbf{x}}(t) & = & \mathbf{A} \mathbf{x}(t) - \mathbf{F}(\bar{\mathbf{x}}(t)) ,\\
 \dot{\bar{x}}(t) & = & \gamma \left( x(t) - \bar{x}(t) \right) - \omega_{0} \bar{y}(t) , \nonumber \\
 \dot{\bar{y}}(t) & = & \gamma \left( y(t) - \bar{y}(t) \right) + \omega_{0} \bar{x}(t) , \nonumber
\end{eqnarray}
where $\bar{x}$ and $\bar{y}$ denote the filtered versions of $x$ and $y$, respectively. Note that the feedback force
$\mathbf{F}(\bar{\mathbf{x}}(t))$ is now generated from the filtered state vector $\mathbf{x}(t)$ which consists of
the two filtered variables $\bar{x}$ and $\bar{y}$:
\begin{equation}
\mathbf{F}(\bar{\mathbf{x}}(t))= K \sum_{n=0}^{\infty} R^{n} \left[ \bar{\mathbf{x}}(t-n \tau) -
\bar{\mathbf{x}}(t-(n+1) \tau) \right].
\end{equation}
Equivalently to the additional differential equations in Eq.~(\ref{eqn:filtered_system}), the filtering, for instance
 in the case of a low pass filter, \textit{i.e.}, $\omega_0=0$, can be expressed by convolution integrals, where the
filtered counterparts of $x(t)$ and $y(t)$ are given as follows:
\begin{eqnarray}
	\bar{x}(t) & = & \gamma \int_{-\infty}^{t}x(t') e^{-\gamma(t-t')} dt'\\
	\bar{y}(t) & = & \gamma \int_{-\infty}^{t}y(t') e^{-\gamma(t-t')} dt'.
\end{eqnarray} 

The system of differential equations~(\ref{eqn:filtered_system}) yields a characteristic equation of the form
\begin{eqnarray}
	\label{eq:char_low}
  \lefteqn{ \pm i \left[ \omega_{0} \left( \lambda -\Lambda \right) + \omega \left( \gamma + \Lambda \right) \right] }
\\
  && =  \gamma K \frac{1-e^{-\Lambda \tau}}{1-R e^{-\Lambda \tau}} + \omega_{0} \omega - \left( \lambda - \Lambda \right) \left( \gamma + \Lambda \right) . \nonumber
\end{eqnarray}

The minimum feedback gain can be calculated in dependence on $\omega_{0}$ and $\gamma$ in similarity to Eq.~(\ref{eq:Kmin}):
\begin{equation}
K_{min}(\omega_{0},\gamma) = \frac{\lambda\left(1+R\right)}{2} \left[ 1 + \frac{\left(\omega_{0}+\omega \right)^{2} + \frac{2 \omega_{0} \omega \lambda}{\gamma} }{\left(\gamma-\lambda \right)^{2}} \right] .
\label{eqn:bandpass_k_min}
\end{equation}

Note that at this special value of the feedback gain $K$, the time delay is no longer given by the intrinsic value $\tau=\pi/\omega$, neither is it the optimal time delay in the general case of nonzero $\omega_{0}$ and finite values of $\gamma$. Adjusting the time delay according to the choice of $\omega_{0}$ and $\gamma$ leads to boundaries of the domain of control that can only be understood in the scope of a parametric representation of the boundaries in the $(K,R)$-plane. Such parametric equations can be derived in analogy to Eq.~(\ref{eq:KRparametric_2}). However, the resulting equations for the case of the filtered system are not shown here due to the complexity of the terms.

\begin{figure}[ht]
\includegraphics[width= \linewidth,angle=0]{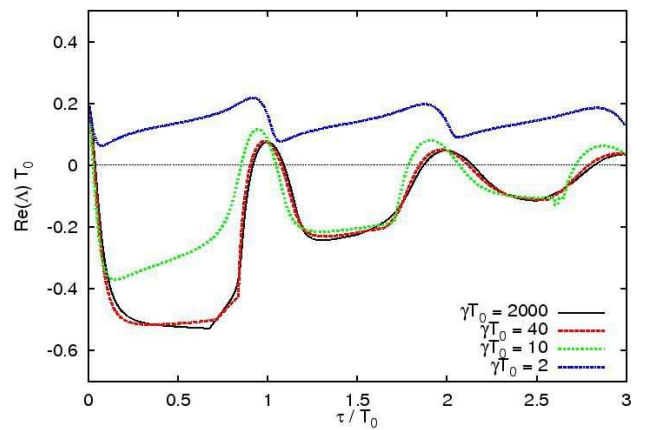}
\caption{\label{fig:ReLambda_gamma}(Color online)  Largest real part of the eigenvalues $\Lambda$ as a function of $\tau$ for
different filter widths $\gamma$ and fixed $\omega_{0}=0$. The black, red, green, and blue curves correspond to $\gamma T_{0}=2000$, $40$, $10$, and $2$, respectively. The other control parameters are fixed as $R=0.7$ and $KT_0=0.6$. The parameters of the system are as in Fig.~\ref{fig:ReLambda}.}
\end{figure}

\begin{figure}[ht]
\includegraphics[width= \linewidth,angle=0]{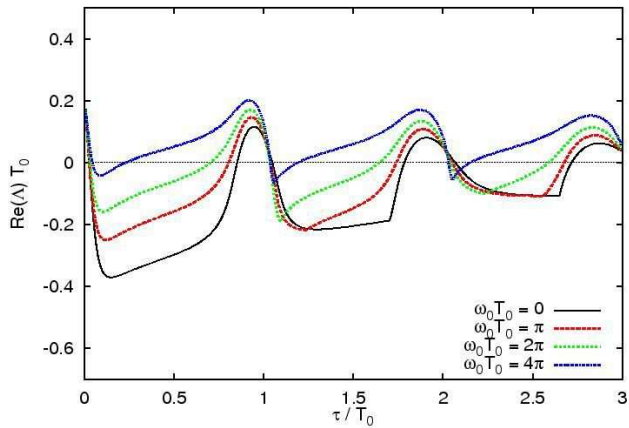}
\caption{\label{fig:ReLambda_omega0}(Color online)  Largest real part of the eigenvalues $\Lambda$ as a function of $\tau$ for
different mean values of the filter $\omega_{0}$ and fixed $\gamma T_{0}=10$. The black, red, green, and blue curves correspond to $\omega_{0} T_{0}=0$, $\pi$, $2\pi$, and $4\pi$, respectively. The other control parameters are fixed as $R=0.7$ and $KT_0=0.6$. The parameters of the system are as in Fig.~\ref{fig:ReLambda}.}
\end{figure}

Figures~\ref{fig:ReLambda_gamma} and \ref{fig:ReLambda_omega0} depicts numerical solutions of the charactistic equation
Eq.~(\ref{eq:char_low}) for different choices of the parameters $\gamma$ and $\omega_{0}$. In
Fig.~\ref{fig:ReLambda_gamma} the largest real part of the eigenvalues is shown in dependence on the time delay for
different values of the filter width $\gamma$ in dependence on the time delay. The parameter $\omega_{0}$ is fixed to
$0$, which correspnds to the case of a low pass filter. The black, red, green, and blue curves correspond to $\gamma
T_{0}=2000$, $40$, $10$, and $2$, respectively. The other control parameters are fixed as $R=0.7$ and $KT_0=0.6$. It can
be seen, that large values of $\gamma$ show similar behavior as in the unfiltered system in Fig.~\ref{fig:ReLambda}.
Decreasing $\gamma$ flattens the curves and shifts them up towards positive real parts. Therefore, no control is
possible for the case of $\gamma T_{0}=2$. Additionally, the optimal time delay is shifted to smaller values, which can
be seen from the case $\gamma T_{0}=10$.

In Fig.~\ref{fig:ReLambda_omega0} similar curves are shown for fixed $\gamma T_{0}=10$ and different choices of the
filter's mean value $\omega_{0}$. The black, red, green, and blue curves correspond to $\omega_{0} T_{0}=0$, $\pi$,
$2\pi$, and $4\pi$, respectively. The other control parameters are fixed as $R=0.7$ and $KT_0=0.6$. It can be observed
that the increase of $\omega_{0}$ shifts the curves further upwards to positive real parts, leading to a less stable
system. The range of the time delay for successful control shrinks.

\begin{figure}[ht]
\includegraphics[width= \linewidth,angle=0]{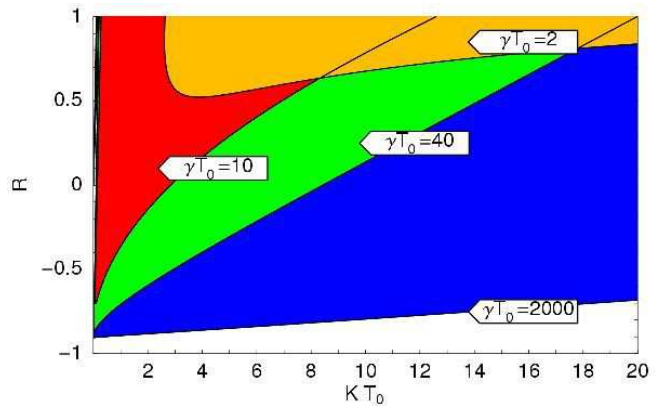}
\caption{\label{fig:KR_gamma}(Color online) Domain of control in the $(K,R)$ plane for different values of the filter width $\gamma$ and fixed $\omega_{0}=0$. The blue, green, red, and yellow areas refer to values of $\gamma T_{0}= 2000$, $40$, $10$, and $2$, respectively. The other control parameters are fixed as $R=0.7$ and $\tau=T_{0}/2$. The parameters of the system are as in Fig.~\ref{fig:ReLambda}.}
\end{figure}

The domain of control can also be investigated in the $(K,R)$ plane. Results are shown in Fig.~\ref{fig:KR_gamma} for
different values of the filter width $\gamma$ and fixed $\omega_{0}=0$. The blue, green, red, and yellow areas refer to
values of $\gamma T_{0}= 2000$, $40$, $10$, and $2$, respectively. The other control parameters are fixed as $R=0.7$ and
$\tau=T_{0}/2$. The domain of control for large values of $\gamma$, here for instance $\gamma T_0=2000$, looks very
similar to that in the case of the unfiltered system, that is depicted in Fig.~\ref{fig:RK}. Decreasing the value of
$\gamma$, the region of control shrinks. The lower right boundary is shifted up towards larger values of $R$. The left
boundary, which corresponds to $K_{min}$, is shifted towards larger values of $K$. This effect is very small for $\gamma
T_{0}=40$ and $10$. For $\gamma T_{0}=2$, the effect is much more pronounced.

\begin{figure}[ht]
\includegraphics[width= \linewidth,angle=0]{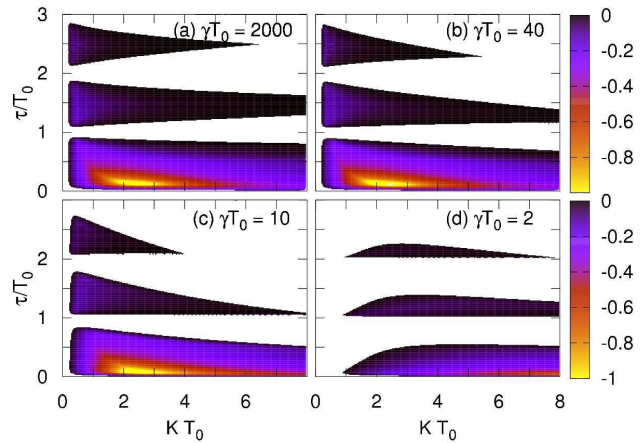}
\caption{\label{fig:Ktau_gamma}(Color online) Domain of control in the $(\tau,K)$ plane for different values of
the filter width $\gamma$ and fixed $\omega_{0}=0$. Panels (a), (b), (c), and (d) correspond to values of $\gamma T_{0}=2000$, $40$, $10$, and $2$, respectively, where the time delay is fixed at $\tau = T_0/2$ and $R=0.7$. The parameters of the system are as in Fig.~\ref{fig:ReLambda}.}
\end{figure}

Figure~\ref{fig:Ktau_gamma} shows the domain of control in the $(K,\tau)$ plane for different values of
the filter width $\gamma$ and fixed $\omega_{0}=0$. Panels (a), (b), (c), and (d) correspond to values of $\gamma T_{0}=2000$, $40$, $10$, and $2$, respectively, where the time delay is fixed at $\tau = T_0/2$ and $R=0.7$. The case of $\gamma T_{0}=2000$ looks very similar to the case of the unfiltered system, as shown in Fig.~\ref{fig:Ktau}. For decreasing value of $\gamma$, the tongues shrink in $K$ direction. The minimum value of $K$ increases and the maximum value becomes smaller. Additionally, the tongues are bent down towards smaller values of the time delay $\tau$ with decreasing $\gamma$. This leads to an optimal $\tau$ that is smaller than in the unfiltered case.

\begin{figure}[ht]
\includegraphics[width= \linewidth,angle=0]{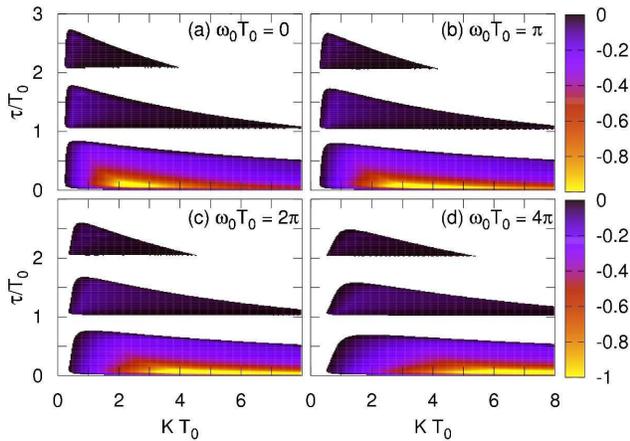}
\caption{\label{fig:Ktau_omega0}(Color online) Domain of control in the $(\tau,K)$  plane for different values of
the filter's mean value $\omega_{0}$ and fixed $\gamma T_{0}=10$. Panels (a), (b), (c), and (d) correspond to values of $\omega_{0} T_{0}=0$, $\pi$, $2\pi$, and $4\pi$, respectively, where the time delay is fixed at $\tau = T_0/2$ and $R=0.7$. The parameters of the system are as in Fig.~\ref{fig:ReLambda}.}
\end{figure}

In Fig.~\ref{fig:Ktau_omega0} the filter width is fixed to $\gamma T_{0}=10$, the domain of control is shown for different values of the filter's mean values $\omega_{0}$. Panels (a), (b), (c), and (d) correspond to values of $\omega_{0} T_{0}=0$, $\pi$, $2\pi$, and $4\pi$, respectively, where the time delay is fixed at $\tau = T_0/2$ and $R=0.7$. The size of the domain of control is only slightly changed with increasing $\omega_{0}$. The domains are flattened on the upper side for larger values of $\omega_{0}$. The region of optimal control, denoted by yellow color code, is additionally shifted slightly towards larger values of the feedback gain $K$ and shrinks in $\tau$ direction. Overall, the variation of $\omega_{0}$ has very little effect on the domain of control in the $(K,\tau)$-plane.

\section{\label{sec:conclusion}Conclusion}
In conclusion, we have shown that extended time-delayed feedback can be used to stabilize unstable steady states of
focus type. By introduction of an additional memory parameter, this method is able to control a larger range of unstable
fixed points compared to the original TDAS scheme. However, the degree of stability, measured by the absolute value of
the real part of the eigenvalue, is generally decreased. We have investigated the domain of control in various one- and
two-dimensional projections of the space spanned by the control parameters. Furthermore, we have discussed also effects
of nondiagonal coupling, nonzero control loop latency, and bandpass filtering of the control signal, which are relevant
in an experimental realization of the ETDAS control method. We found that a proper adjustment of the time delay is
able to compensate for reducing stabilization abilities of the control method, for instance, due to latency or feedback
phase.

We point out that the results obtained in this paper are accessible for applications in the context of all-optical control of intensity oscillations of semiconductor lasers as investigated in \cite{SCH06a}.

\section{Acknowledgements}
This work was supported by Deutsche Forschungsgemeinschaft in the framework of Sfb 555.


\end{document}